\documentclass[%
preprint,
 amsmath,amssymb,
aps,
]{revtex4-1}

\usepackage{graphicx}
\usepackage{dcolumn}
\usepackage{bm}
\usepackage{MnSymbol}

\begin{document}%


\title{Room-temperature spin transport in $n$-Ge probed by four-terminal nonlocal measurements}

\author{Michihiro Yamada,$^{1}$\thanks{E-mail address: michihiro@ee.es.osaka-u.ac.jp} Makoto Tsukahara,$^{1}$ Yuichi Fujita,$^{1}$ Takahiro Naito,$^{1}$ Shinya Yamada,$^{1,2}$ Kentarou Sawano,$^{3}$ and Kohei Hamaya$^{1,2}$\thanks{E-mail address: hamaya@ee.es.osaka-u.ac.jp}}
\affiliation{
$^{1}$Graduate School of Engineering Science, Osaka University, 1-3 Machikaneyama, Toyonaka 560-8531, Japan}
\affiliation{
$^{2}$Center for Spintronics Research Network, Osaka University, 1-3 Machikaneyama, Toyonaka 560-8531, Japan}
\affiliation{
$^{3}$Advanced Research Laboratories, Tokyo City University, 8-15-1 Todoroki, Tokyo 158-0082, Japan.}

\date{\today}

\begin{abstract}
We demonsrtate electrical spin injection and detection in {\it n}-type Ge ({\it n}-Ge) at room temperature using four-terminal nonlocal spin-valve and Hanle-effect measurements in lateral spin-valve (LSV) devices with Heusler-alloy Schottky tunnel contacts. 
The spin diffusion length ($\lambda$$_{\rm Ge}$) of the Ge layer used ($n \sim$ 1 $\times$ 10$^{19}$ cm$^{-3}$) at 296 K is estimated to be $\sim$ 0.44 $\pm$ 0.02 $\mu$m. 
Room-temperature spin signals can be observed reproducibly at the low bias voltage range ($\le$ 0.7 V) for LSVs with relatively low resistance-area product ($RA$) values ($\le$ 1 k$\Omega$$\mu$m$^{2}$). 
This means that the Schottky tunnel contacts used here are more suitable than ferromagnet/MgO tunnel contacts ($RA \ge$ 100 k$\Omega$$\mu$m$^{2}$) for developing Ge spintronic applications.
\end{abstract}


\maketitle
Owing to the future intrinsic limits of downsizing of silicon-based conventional complementary metal-oxide-semiconductor (CMOS) transistors, novel devices with additional functionalities should be developed. 
Spin-based electronics (spintronics) is expected to enhance device performances because of its nonvolatility, reconstructibility, low power consumption, and so forth.\cite{Igor,Hirohata,Taniyama,Yuasa,Tanaka} 
To introduce the use of spintronics into the Si-based semiconductor industry, it will become important to explore spintronic technologies compatible with Si.\cite{Jansen1,Appelbaum,Jonker,Si_HamaGr,Suzuki,Ishikawa} 
In recent years, germanium has been attracting much attention as a channel material for next-generation CMOS transistors because its electron and hole mobility are twice and four times as large as those in Si, respectively.\cite{Kuzum,Zhang1} 

In line with this research, there have been many studies to date on the development of spintronic technologies using Ge.\cite{Zhou,KAIST,Ge_Hama1,Ge_Hama2,Ge_Hama3,AIST,Jamet} 
For heavily doped {\it n}-Ge (10$^{18}$ cm$^{-3}$$\le n \le$ 10$^{19}$ cm$^{-3}$), the spin diffusion length ($\lambda$$_{\rm Ge}$) at room temperature, one of the key parameters in Ge spintronics, has been reported by some experimental methods. 
First, the value of $\lambda$$_{\rm Ge}$ at room temperature was estimated to be 0.68 $\mu$m from three-terminal Hanle-effect measurements of CoFe/MgO/{\it n}-Ge devices.\cite{KAIST} 
Next, by analyzing the detected inverse-spin-Hall voltage in Py/{\it n}-Ge/Pd devices, a room-temperature $\lambda$$_{\rm Ge}$ value of 0.66 $\mu$m was expected,\cite{Dushenko_PRL} which is similar to that in Ref. \cite{KAIST}. 
However, there is no report on the room-temperature $\lambda$$_{\rm Ge}$ estimated from four-terminal nonlocal spin-valve and Hanle-effect measurements although this is the most reliable method for exploring spin transport in nonmagnets.\cite{Johnson, Jedema, Kimura, Lou_NatPhys} 


Here, we report on room-temperature four-terminal nonlocal spin signals and Hanle-effect curves obtained under parallel and anti-parallel magnetization configurations of Co$_{2}$FeAl (CFA) Schottky tunnel contacts in {\it n}-Ge-based lateral spin-valves (LSVs). 
This means that generation, manipulation, and detection of pure spin currents in {\it n}-Ge are reliably demonstrated at room temperature by all-electrical means for the first time. 
The $\lambda$$_{\rm Ge}$ value of the {\it n}-Ge layer used here ($n \sim$ 1 $\times$ 10$^{19}$ cm$^{-3}$) is estimated to be $\sim$ 0.44 $\pm$ 0.02 $\mu$m.
Room-temperature spin signals can be observed for LSVs with relatively low resistance-area product ($RA$) values ($\le$ 1 k$\Omega$$\mu$m$^{2}$), which are much lower than that of ferromagnet/MgO tunnel contacts in the low bias voltage range in room-temperature Ge spintronics.\cite{KAIST,Jamet} This study will pave a way to develop spintronic applications with Ge technologies.

First, the growth of Ge layers used in this study for spin transport is explained. 
We formed an undoped Ge(111) layer ($\sim$28 nm) grown at 350 $^\circ$C (LT-Ge) on a commercial undoped Si(111) substrate ($\rho$ $\sim$ 1000 $\Omega$cm), followed by an undoped Ge(111) layer ($\sim$70 nm) grown at 700 $^\circ$C (HT-Ge).\cite{Sawano_TSF}   
As the spin transport layer, we grew a 140-nm-thick phosphorous (P)-doped $n$-Ge(111) layer (doping concentration $\sim$ 10$^{19}$ cm$^{-3}$) by molecular beam epitaxy (MBE) at 350 $^\circ$C on top of the HT-Ge layer. 
The carrier concentration ($n$) in the $n$-Ge(111) layer was estimated to be $n \sim$ 1 $\times$ 10$^{19}$ cm$^{-3}$. 
Because the HT-Ge layer on LT-Ge/Si(111) has $p$-type conduction and a relatively high resistivity (high spin resistance) compared to the spin transport ($n$-Ge) layer, we can ignore the spin diffusion into the HT-Ge layer. 
To promote the tunneling conduction of electron spins through the Schottky barriers,\cite{Ge_Hama2,Ge_Hama1,Ge_Hama3} a P $\delta$-doped Ge layer ($n^{+}$-Ge) with an ultra thin Si insertion layer was grown on top of the spin-transport layer.\cite{MYamada} 
As the spin injector/detector, we used Co$_{2}$FeAl (CFA) grown by low-temperature MBE because it is relatively simple to grow compared to Co$_{2}$FeAl$_{0.5}$Si$_{0.5}$, which was previously reported in Ref. \cite{Ge_Hama3}. Here CFA is also expected to exhibit relatively high spin polarization.\cite{Sukegawa} Detailed growth procedures have been published elsewhere.\cite{SYamada}  
\begin{figure}[t]
\begin{center}
\includegraphics[width=8 cm]{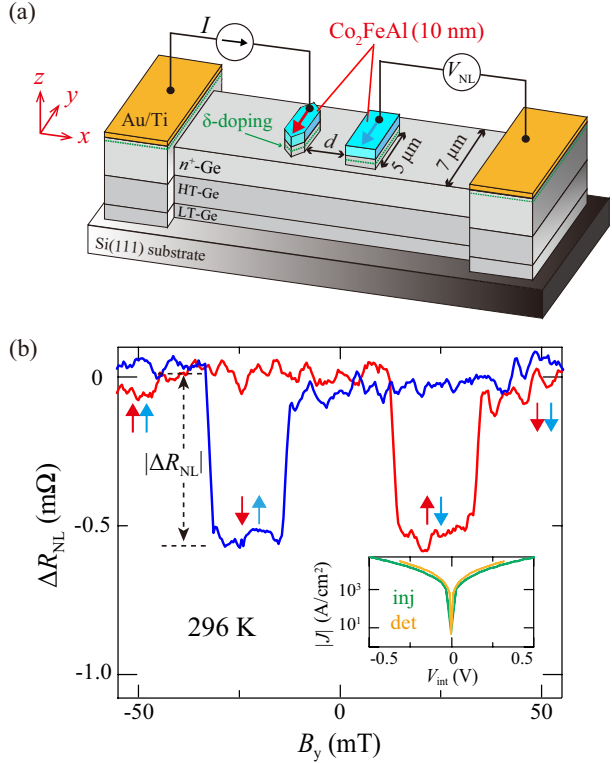}
\caption{(Color online) (a) Schematic diagram of a lateral four-terminal device with Co$_{2}$FeAl/{\it n}$^{+}$-Ge contacts. (b) Four-terminal nonlocal magnetoresistance curve at 296 K. Inset shows $J-V$ curves of the Co$_{2}$FeAl/{\it n}$^{+}$-Ge spin injector and detector at 296 K. }
\end{center}
\end{figure}

Figure 1(a) shows a schematic illustration of the fabricated LSVs for four-terminal nonlocal measurements.\cite{Johnson, Jedema, Kimura} 
Two different CFA/$n$-Ge contacts with 0.4 $\times$ 5.0 $\mu$m$^{2}$ and 1.0 $\times$ 5.0 $\mu$m$^{2}$ in size were fabricated by conventional electron beam lithography and Ar-ion milling. 
The edge-to-edge distances $d$ between the CFA/$n^{+}$-Ge contacts were measured to be 0.24 to 0.55 $\mu$m. 
Representative current-voltage characteristics ($J-V$ curves) of the fabricated CFA/$n^{+}$-Ge Schottky tunnel contacts are shown in the inset of Fig. 1(b). 
Because there is no rectifying behavior, we can judge that the tunneling conduction of electrons through the CFA/$n$-Ge interfaces is demonstrated irrespective of the influence of the strong Fermi level pinning at the metal/$n$-Ge interface.\cite{Dimoulas} 
By applying in-plane magnetic fields ($B$$\rm_{y}$), four-terminal nonlocal magnetoresistance ($\Delta$$R$$_{\rm NL}$ $=$ $\Delta$$V$$_{\rm NL}$/$I$) curves are measured at $I$ = -2 mA at room temperature (296 K), as shown in the main panel of Fig. 1(b). Here the negative sign of $I$ ($I$ $<$ 0) means that spin-polarized electrons are injected from CFA into $n$-Ge. 
Evident hysteretic behavior of $\Delta R_{\rm NL}$, depending on the magnetization configuration between the two CFA contacts, can be observed even at room temperature. 
These features can be reproducibly observed in many LSVs, as shown in the discussion below. 
For the presented LSV ($d =$ 0.24 $\mu$m), the amplitude of the spin signal, $|\Delta$$R$$_{\rm NL}|$, is approximately 0.54 m$\Omega$ at room temperature.
\begin{figure}[b]
\begin{center}
\includegraphics[width= 7.5 cm]{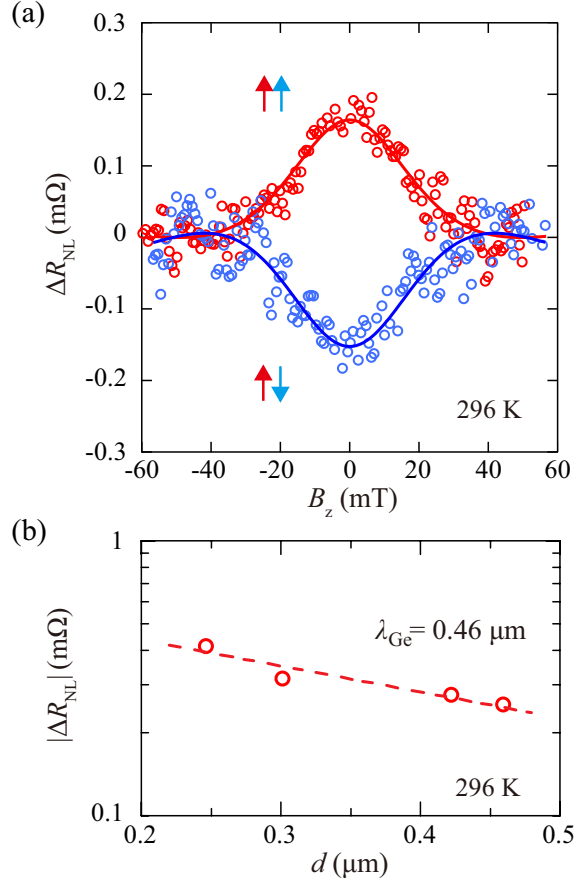}
\caption{(Color online) (a) Four-terminal nonlocal Hanle effect curves of a CFA/$n^{+}$-Ge LSV for parallel and antiparallel magnetization configurations at 296 K at $I$ $=$ $-$2.0 mA.
(b) Plot of $|\Delta R_{\rm NL}|$ versus $d$ at 296 K. Dashed line is a result fitted to ${\frac{|{P_{\rm inj}|}|{P_{\rm det}|}{\rho_{\rm Ge}}{\lambda_{\rm Ge}}}{S}}{\exp\left(-\frac{d}{\lambda_{\rm Ge}}\right)}$. }
\end{center}
\end{figure}

By applying out-of-plane magnetic fields ($B$$\rm_{z}$), we also measure four-terminal nonlocal Hanle-effect curves under parallel and anti-parallel magnetization states between the CFA electrodes. 
In Fig. 2(a), we can see clearly nonlocal Hanle-effect curves, indicating precession of the pure spin currents in the {\it n}-Ge layer at room temperature, for both the parallel and antiparallel magnetization states in an LSV having $d =$ 0.5 $\mu$m.  
This is the first observation of four-terminal nonlocal Hanle-effect curves at room temperature in {\it n}-Ge, that were detected entirely by electrical means.
According to the one-dimensional spin drift diffusion model,\cite{Lou_NatPhys,Jedema} the four-terminal nonlocal Hanle-effect curves can be expressed as follows. 
\begin{equation}
\Delta R_{\rm NL}(B_{\rm z}) = \pm A{ {\int_0^{\infty}}{\phi(t)}{\rm cos}({\omega}_{L}t){\exp\left(-\frac{t}{\tau_{\rm Ge}}\right)}dt},
\end{equation}
where $A =$ ${\frac{{P_{\rm inj}}{P_{\rm det}}{\rho_{\rm Ge}}D}{S}}$, $\phi(t) =$ $\frac{1}{\sqrt{4{\pi}Dt}}{\exp\left(-\frac{L^{2}}{4Dt}\right)}$, $\omega$$_{L}$ (= $g$$\mu$$_{\rm B}$$B$$_{z}$/$\hbar$) is the Larmor frequency, $g$ is the electron $g$-factor ($g$ = 1.56) of Ge,\cite{Vrijen} $\mu$$_{\rm B}$ is the Bohr magneton, $P_{\rm inj}$ and $P_{\rm det}$ are the spin polarizations of electrons in {\it n}-Ge, $\rho_{\rm Ge}$ is the resistivity ($\rho$$_{\rm Ge} \sim$ 1.9 m$\Omega$cm), $S$ is the cross section ($S =$ 0.98 $\mu$m$^{2}$) of the {\it n}-Ge layer used here, and $L$ is the center-to-center distance between the spin injector and detector ($L =$ 1.2 $\mu$m). 
The solid curves presented in Fig. 2(a) indicate the results fitted to Eq.(1). 
In consequence, the $\tau$$_{\rm Ge}$ and $D$ values are determined to be 0.25 ns and 7.2 cm$^{2}$/s, respectively.
Here the influence of the contact-induced spin relaxation on the Hanle analysis of our LSVs can be ignored because the $\tau$$_{\rm Ge}$ value obtained using Eq. (5) in Ref. \cite{Brien_PRB} was nearly equivalent to that obtained using Eq. (1). 
In addition, according to the relation $\lambda_{\rm Ge}$ $=$ $\sqrt{D\tau_{\rm s}}$, we can calculate a $\lambda_{\rm Ge}$ value of 0.42 $\mu$m at room temperature for the $n$-Ge layer used here. 
In Fig. 2(b), we also investigate the $d$ dependence of $|\Delta$$R$$_{\rm NL}|$ for evaluating $\lambda_{\rm Ge}$ using the following equation, 
$|\Delta R_{\rm NL}|$ $=$ ${\frac{|{P_{\rm inj}|}|{P_{\rm det}|}{\rho_{\rm Ge}}{\lambda_{\rm Ge}}}{S}}{\exp\left(-\frac{d}{\lambda_{\rm Ge}}\right)}$. 
Here we use LSVs other than that shown in Fig. 2(a) because we should confirm the reliability of the $\lambda_{\rm Ge}$ value estimated from the same$n$-Ge layer. 
As a result, we can obtain a $\lambda_{\rm Ge}$ value of 0.46 $\mu$m. 
From both the Hanle effect and $d$ dependence analyses, we regard $\lambda_{\rm Ge}$ at room temperature as 0.44 $\pm$ 0.02 $\mu$m, which is slightly different from the values obtained in previous room-temperature works.\cite{KAIST,Dushenko_PRL} 

Recently, by utilizing a Co-based Heusler alloy as a spin injector and detector, we have obtained large and reliable spin signals at low temperatures and gained insight into the spin relaxation mechanism in {\it n}-Ge from low temperature to near room temperature.\cite{Ge_Hama3} 
Theoretical analyses revealed that the spin relaxation mechanism in {\it n}-Ge (10$^{18}$ cm$^{-3}$$\le n \le$ 10$^{19}$ cm$^{-3}$) at low temperatures is dominated by donor-driven intervalley spin-flip scattering.\cite{Ge_Hama3,Song_PRL} 
It should be noted that, however, at room temperature, the phonon-induced intervalley spin-flip scattering cannot be ignored.\cite{Ge_Hama3} 
Considering both contributions to $\tau$$_{\rm Ge}$, we can theoretically calculate a room temperature $\tau$$_{\rm Ge}$ value of $\sim$ 0.2 ns,\cite{Ge_Hama3} which leads to a $\lambda_{\rm Ge}$ of 0.45 $\mu$m.
The experimentally obtained data in this study ($\lambda_{\rm Ge} =$ 0.44 $\pm$ 0.02 $\mu$m) are consistent with the above theoretical $\lambda_{\rm Ge}$ value. 
In terms of the theoretical spin relaxation mechanism,\cite{Song_PRL} the estimated $\lambda_{\rm Ge}$ value in this study is more reliable than those estimated by other methods.\cite{KAIST,Dushenko_PRL}  

A comparison of $\lambda_{\rm Ge}$ and $\lambda_{\rm Si}$ at room temperature shows that the obtained $\lambda_{\rm Ge}$ is obviously smaller than the reported $\lambda_{\rm Si}$ ( $\sim$ 1.0 $\mu$m).\cite{Ishikawa} 
For both Ge and Si, although the spin relaxation mechanism is dominated by the donor-driven and phonon-induced intervalley spin-flip scattering predicted theoretically\cite{Song_PRL} and experimentally,\cite{Ge_Hama3,Ishikawa} the contributions of the Bohr radius and spin-orbit coupling in Ge to the spin scattering rate are relatively large compared to those in Si.
Therefore, reducing the donor concentration in {\it n}-Ge might be important to enhance $\lambda_{\rm Ge}$ at room temperature in future. 

\begin{figure}[t]
\begin{center}
\includegraphics[width= 7.5 cm]{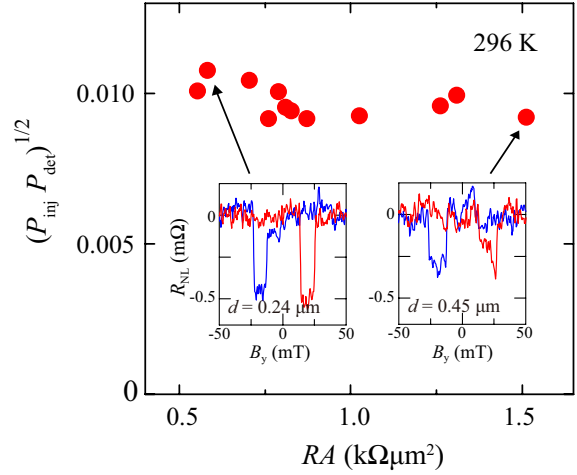}
\caption{(Color online) Summary of $\sqrt{P_{\rm inj}P_{\rm det}}$ versus $RA$ at 296 K for various LSV devices with Co$_{2}$FeAl/{\it n}$^{+}$-Ge contacts. Insets show representative spin signals for LSVs with $RA$ values of $\sim$0.5 k$\Omega$$\mu$m$^{2}$ and $\sim$1.5 k$\Omega$$\mu$m$^{2}$.}
\end{center}
\end{figure}

From the results shown in Fig. 2, the spin polarization created in {\it n}-Ge can be expected. 
Figure 3 shows a plot of room-temperature spin polarization, $\sqrt{P_{\rm inj}P_{\rm det}}$, versus the $RA$ value in the low bias voltage range ($\le$ 1.0 V) for various LSV devices. We cannot see the effect of the $RA$ value of the used contacts on the spin polarization created in {\it n}-Ge.  
Here, for LSVs with $RA$ $\sim$ 1 k$\Omega$$\mu$m$^{2}$, the $|\Delta R_{\rm NL}|$ values were reliably observable, and this range was much lower than that in previous room-temperature works.\cite{KAIST,Jamet} 
We note that the use of our Schottky tunnel barriers with inserted $\delta$-doped layers\cite{Ge_Hama3} enables us to simultaneously demonstrate room-temperature spin transport and low $RA$ spin injection/detection. 
This technique will open a way to develop Ge spintronic applications with low power consumption. We will comment on this point later. 
Unfortunately, the spin polarization created in {\it n}-Ge is still small, $\sqrt{P_{\rm inj}P_{\rm det}} \sim$ 0.01, as shown in Fig. 3.
According to the conventional spin diffusion theory,\cite{Fert} the ratio of $RA$ ($=$ $r_{\rm b}$) to $r_{\rm N}$ is important for obtaining a large the spin signal in an LSV with tunnel barriers, where $r_{\rm N}$ = $\rho_{\rm Ge}$ $\times$ $\lambda_{\rm Ge}$. 
Here, because $r_{\rm N}$ is estimated to be $\sim$ 8.4 $\Omega$$\mu$m$^{2}$ in this study, $RA$/$r_{\rm N}$ is $\sim$ 10$^{2}$, which is not an optimum condition.\cite{Fert} 
To enhance the spin signals in our LSVs, it might be important to control the $RA$ value toward $\sim$10 $\Omega$$\mu$m$^{2}$ by engineering the ferromagnet/{\it n}$^{+}$-Ge junctions. 
However, we have to consider the spin absorption effect at the low-$RA$ interface in future. 
Although the bulk spin polarization of CFA electrodes is expected to be high,\cite{Sukegawa} the quality of the CFA electrode near the heterointerface may not be sufficient to achieve device applications. 
If the bulk spin polarization of the low-temperature grown CFA near the interface is degraded,\cite{Lazarov} we should improve the spin polarization created in {\it n}-Ge by achieving high-quality ordered structures. 
In the study of GaAs-based LSVs, the spin injection/detection efficiency has reportedly be improved by using high-quality Co-based Heusler alloys.\cite{Bruski,Crowell,Tezuka,Uemura_PRBR}  
To obtain even higher spin signals at room temperature in Ge-based LSVs, we should further improve the quality of the Co-based Heusler alloys on Ge.

Finally, we comment on the technical advantage of using Schottky tunnel barriers for Ge spintronic applications. 
According to our previous works,\cite{Hamaya_FLP} high-quality Heulser-alloy/Ge heterointerfaces grown by low-temperature MBE can reduce the influence of the strong Fermi level pinning on the Schottky barrier height.
For such Heulser-alloy/{\it n}-Ge junctions, we can apply a developed P $\delta$-doping technique\cite{MYamada} to reduce the width of the Schottky barrier. 
By utilizing both methods for Schottky tunnel junctions, we can realize high-quality Heulser-alloy/Ge contacts with relatively low $RA$ values ($\le$ 1 k$\Omega$$\mu$m$^{2}$).\cite{MYamada,Ge_Hama3} 
For ferromagnet/MgO/{\it n}-Ge junctions, on the other hand, there are many reports on the use of the barrier layers (1 $-$ 2 nm) to obtain relatively high $RA$ values ($\ge$ 100 k$\Omega$$\mu$m$^{2}$) in the low bias voltage range ($\le$ 1.0 V). 
In addition, complete Fermi level depinning cannot be observed yet by inserting crystalline MgO barriers between bcc-ferromagnets and {\it n}-Ge.\cite{Zhou_FLP} 
For these reasons, our Schottky tunnel barriers are more effective than ferromagnet/MgO barriers for simultaneously demonstrating room-temperature spin transport and low $RA$ contacts in the Ge spintronic devices. 

In summary, we reported on room-temperature four-terminal nonlocal spin signals and Hanle-effect curves obtained under parallel and anti-parallel magnetization states of Co$_{2}$FeAl (CFA) Schottky tunnel contacts in {\it n}-Ge-based LSVs. 
Using the analyses of the Hanle-effect curves and $d$-dependent nonlocal spin signals, we extracted a room temperature $\lambda$$_{\rm Ge}$ value of $\sim$ 0.44 $\pm$ 0.02 $\mu$m for {\it n}-Ge ($n \sim$ 1 $\times$ 10$^{19}$ cm$^{-3}$). 
We note that room-temperature spin signals can be observed for LSVs with a relatively low $RA$ of $\sim$ 1 k$\Omega$$\mu$m$^{2}$, which is much lower than that of ferromagnet/MgO tunnel contacts obtained in previous room-temperature works.\cite{Jamet,KAIST} This study will pave a way to develop Ge spintronic applications with low power consumption.

This work was partially supported by a Grant-in-Aid for Scientific Research (A) (No. 16H02333) from the Japan Society for the Promotion of Science (JSPS),  and a Grant-in-Aid for Scientific Research on Innovative Areas `Nano Spin Conversion Science' (No. 26103003) from the Ministry of Education, Culture, Sports, Science and Technology (MEXT). M.Y. acknowledges scholarships from the Toyota Physical and Chemical Research Institute Foundation. Y.F. acknowledges JSPS Research Fellowships for Young Scientists.

\end{document}